\documentclass[11pt,twoside]{article}
\usepackage{asp2010}

\resetcounters

\usepackage{epsfig}
\begin{document}

\title{Physics of stars understood/expected from asteroseismology}
\author{Hideyuki Saio
%\affil{$^1$Institution Full Address for Author1}
%\affil{$^2$Institution Full Address for Author2}
\affil{Astronomical Institute, Graduate School of Science, Tohoku University, Sendai 980-8578, Japan}}

\begin{abstract}
What can be learned about the physics of stellar interiors from studying stellar oscillations?
This review address the potential to improve our understandings of  convective core overshoot and of more general convection-related effects, microscopic diffusion,  rotation and magnetic fields, and finally evolution-induced period changes.
\end{abstract}

\section{Introduction}
The term `asteroseismology' appeared in the title of a published paper for the first time in ``What will  Asteroseismology teach us?" by \citet{jcd84}.  
After that, the number of papers related with asteroseismology has increased exponentially. Based on recent successes in observational techniques, we can safely expect the trend to   even accelerate in the foreseeable future.

A narrow definition of asteroseismology might be to deduce physical parameters and  interior structure from several observed oscillation frequencies; i.e., an extension of Helioseismology to stars. 
Here we are going to adopt a somewhat more relaxed attitude; whenever we can learn \emph{any} stellar physics from the observed oscillations, we will call this asteroseismology.
This paper overviews some of the topics in stellar oscillations; more thorough discussions can be found e.g. in the books by \citet{cox80},\citet{unn89}, or \citet{aer10}.

Everyone in this research field is aware of the very rapid progress of asteroseismology driven mostly by the high-quality space photometries from the MOST, CoRoT, and Kepler missions,  organized ground based complementary photometric and spectroscopic observations, and many theoretical works.
Due to the huge body of results,
I am afraid that I will inevitably miss some important results. I apologize in advance for these blind spots. 

\section{Overshooting from Convective Core}
Convective core overshooting and (rotational) slow mixing in radiative layers are important processes whose strengths are hardly constrained by purely theoretical considerations, but they affect significantly the evolution of stars.
Since the state of the stellar surface gives us little information on the interior mixing, asteroseismology is  anticipated to reveal the extent of these mixing processes.

\subsection{$\beta$ Cephei stars}
$\beta$ Cephei stars are relatively massive ($\gtrsim 8M_\odot$) main-sequence stars with a well developed convective cores; they oscillate in low-oder p/g-modes with periods of a few hours.
In most cases, the number of frequencies is low, and they rotate relatively slowly so that the rotational frequency splittings can be treated perturbatively.
In addition,  surface degrees of the oscillation modes for the observed frequencies can be estimated from their amplitudes at various wavelengths \citep[e.g.,][]{der04,das10}  so that mode identification is not very difficult.
Thus, $\beta$ Cephei stars are well suited for asteroseismology to obtain information on the extent of convective-core overshooting.

The slowly rotating star $\nu$ Eridani is the most intensively studied $\beta$ Cephei variable.
The organized photometric and spectroscopic multisite observations
\citep{han04,aer04,jer05} revealed that the star has, at least, three rotational triplets and 
five independent frequencies including two high-order (SPB-type) g-modes.
Although many detailed seismic investigations for $\nu$ Eri have been performed, some difficulties remain in frequency-fittings and modeling excitation of observed frequencies. 
Table~\ref{tab:betcep} summarizes some of the results. 
Although fitting the frequencies of  the radial mode ($\ell=0, p_1$) and the two triplets ($\ell=1$, $g_1, p_1$) with models imposes no stringent constraint to the core-overshooting parameters, models with $\alpha_{\rm ov} \sim 0.2$ seem to work better in fitting the other frequencies. However, models with a considerable core-overshooting tend to have
lower effective temperature than the photometrically estimated range. 

Table~\ref{tab:betcep} shows that analyses for other $\beta$ Cephei stars also point to  overshooting parameters in the range of $0.1 - 0.4$.
Although there are still considerable differences in the parameters obtained by different authors, most of the analyses consistently required considerable core overshooting in $\beta$ Cephei stars. 
This is, however,  at odds with the SPB stars as discussed below.
 
\begin{table}
\caption{Parameters for some $\beta$ Cephei stars obtained from asteroseismology}
\label{tab:betcep}
\begin{tabular}{@{}lccccccc}
\hline
Name & $M$  & Z &  $X_{\rm c}$ & $\alpha_{\rm ov} $ & $\Omega_{\rm core}/\Omega_{\rm env}$ & note\\
\hline
$\nu$ Eri & $9.0-9.9$  & .015 & $.24-.25$& $0.0 - 0.12$ & $\sim 5$  & 1 \\
               & $8.4 $  & $ .0115$ & $-$ & $0.21$ & $-$  & 2 \\
               & $7.13$ & $.019$ & $.13-.16$ & $0.24 - 0.28$ & $1 - 2.6$ & 3\\
               & $8.8-9.6$  & $.016-.019$ & $-$ & $0.0 - 0.2$ & $5.4-5.8$  &4 \\
               & $9.05$  & $.014$ & $.26$ & $0.17$& $-$  & 5 \\
               & $8.59$  & $.016$ & $.27$ & $0.23$ & $-$  & 5 \\
\hline
$\theta$ Oph & $ 8.2\pm0.3$  & $.009-.015$ & $.38\pm.02$ & $0.44\pm0.07$ & $1- $ & 6\\
      & 9.5 & .02 & .32 &$0.28\pm0.1$ & & 7 \\
\hline
HD129929 & $9.35$  &  $.019$ & .353 & $0.10$ & 3 & 8\\
$\beta$ CMa & $13.5\pm0.5$&   $.019\pm.003$ &  $.126\pm.003$ & $0.20\pm0.05$ & &9 \\
$\delta$ Cet & $10.2\pm0.2$&  .02  &  .25 &  $0.2\pm 0.05$ & & 10\\
\hline
\end{tabular}
1=\citet{pam04}, OPAL, 3 freq.-fit; ~
2=\citet{aus04}, OPAL, 4 freq.-fit, 4Fe$_\odot$; ~  
3=\citet{sua09}, OPAL, $X_0=0.5$;~
4=\citet{dzi08}, OPAL \& OP; ~
5=\citet{das10}, OPAL2 \& OP2;~
6=\citet{bri07}; ~
7=\citet{lov10}, 2D; ~
8=\citet{dup04}; ~
9=\citet{maz06}; ~
10=\citet{aer06}
\end{table}

\subsection{No overshooting needed in SPB stars} 
Slowly pulsating B (SPB) stars are multi-periodic pulsators with periods of 1 to 2 days.  
They are main-sequence stars with masses in a range of $3\lesssim M/M_\odot \lesssim 8$, and  the majority is slowly rotating. 
High radial order g-modes are excited in these stars by the kappa-mechanism works at the `Fe-bump'  ($T\sim2\times10^5$K) as for the low order p/g-modes in $\beta$ Cephei stars.  
Long-period g-modes are excited because the effective temperatures of SPB stars are lower than of the $\beta$ Cephei stars so that the opacity bump is located slightly deeper in the envelope where the thermal timescale is longer.

\begin{figure}
\begin{center}
\epsfig{file=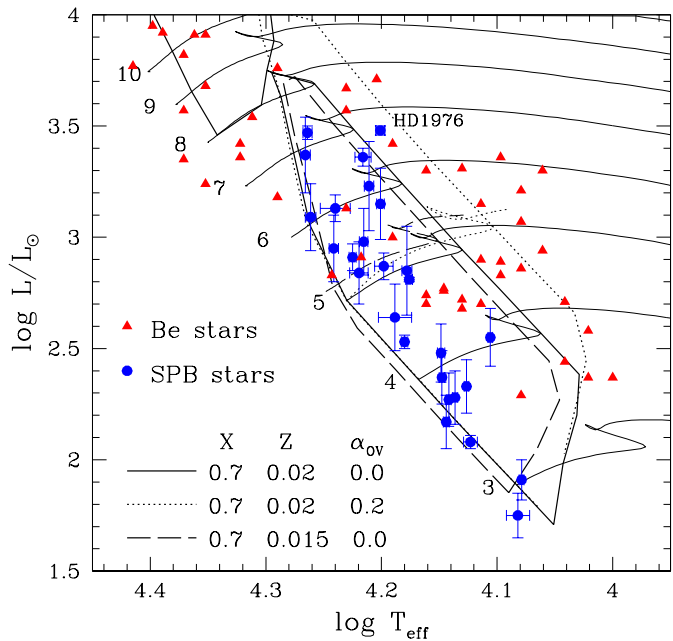,width=0.7\textwidth}
\end{center}
\caption{Evolutionary tracks and instability ranges for g-modes in models with different compositions and with/without core overshooting. The lower part of the instability range for p-modes is also shown. OPAL opacity tables \citep{opal96} were used.   
Parameters for SPB stars are adopted from \citet{dec02,nie02,aer05}, 
and for Be stars from \citet{fre06} 
}
\label{fig_hrdspb}
\end{figure}

The excitation of g-modes is affected not only by the effective temperature but also the internal structure of the stars. 
The red boundary of the  SPB  instability region in the HR diagram coincides with the disappearance of the convective core at the termination of the main-sequence stage. 
In the evolved radiative core, the Brunt-V\"ais\"al\"a frequency  becomes very large so that the wavelengths of the spatial oscillations of a high-order g-mode becomes very short, which in turn enhances the radiative dissipation to damp the g-mode \citep[e.g.][]{gau93}. 
Therefore, the red boundary of the distribution of SPB stars in the HR diagram is a good indicator of the location of the termination of the main-sequence stage which depends on the extent of core-overshooting.
In other words, by comparing the observed red-boundary of the SPB stars with theoretical boundaries we can estimate the extent of core-overshooting in these stars. 
Figure~\ref{fig_hrdspb} compares the distribution of SPB stars (and Be stars) on the HR diagram with the theoretical instability regions for g-modes obtained with/without core-overshooting.
All (but HD 1976) SPB stars lay within the instability region for models without core-overshooting.  
This means that in contrast to the results for $\beta$ Cephei stars no core-overshooting is needed to understand  SPB stars. 
The star HD 1976 is a relatively rapid rotator with  $V\sin i = 125$~km/s; so that rotation might enhance mixing around the convective core.

Figure~\ref{fig_hrdspb} shows a clear difference in the distribution between Be stars, which rotate nearly critically, and  SPB stars, the majority of which rotate slowly.  
Be stars having effective temperatures similar to those of SPB stars show g-mode oscillations \citep{wal05,sai06}, which indicates that these stars should have a convective core and still be in the main-sequence stage. 
To explain the distribution in Fig.~\ref{fig_hrdspb} and the presence of g-mode oscillations  in the intermediate mass ($3\lesssim M/M_\odot \lesssim 7$)  Be stars, extensive core-overshooting, as large as $\sim 0.4H_p$, is needed \citep{lov10be}, in stark contrast to the case of SPB stars.
This may indicate that in rapidly rotating stars  rotational mixing around the convective-core carries extra hydrogen into the convective core mimicking  extensive overshooting from the convective-core boundary, although the pure core-overshooting itself is weak without rotation effects.
More accurate and detailed seismological analyses for $\beta$ Cephei and Be stars are needed to clarify the relation between stellar rotation and mixing in radiative layers around convective cores.

\section{Effects of microscopic diffusion}
In a quiet stellar envelope microscopic diffusion occurs; i.e., each ion sinks or rises depending on the relative strength of  gravity and upward radiation forces. 
Strong microscopic diffusion can affect the excitation/damping of stellar oscillations.
In this section, we discuss briefly such effects in Am stars, sdB stars, and hot pre-white dwarfs.

\subsection{Am stars} \label{sec_am}
In Am (A metallic line) stars, some metals are levitated to the surface due to radiation force and helium is depleted due to gravitational settling.  
Before the era of accurate space photometry it was thought that majority of Am stars lying in the classical instability strip do not pulsate because helium is drained from the second helium ionization zone due to the gravitational settling, which quenched the kappa-mechanism of excitation.  
This view has been overturned by the recent observations by Kepler \citep{bal11} and the ground-based survey WASP with automated telescopes \citep{sma11}. They have found that the majority of Am stars do pulsate and their distribution in the HR diagram is the same as that of $\delta$ Sct variables, although the pulsation amplitudes in Am stars tend to be smaller \citep{sma11}.  
This finding does not agree with the theoretical prediction obtained from models including microscopic diffusions \citep{tur00}; the diffusion in Am stars seems not as strong as predicted by the models.

\subsection{Subdwarf B (sdB) stars} \label{sec:spb}
The sdB stars are hot $(25,000~{\rm K} \lesssim T_{\rm eff} \lesssim 40,000 ~{\rm K})$ and subluminous $(10 \lesssim L/L_\odot \lesssim 50)$ evolved compact ($5\lesssim \log g \lesssim 6$) stars \citep[see e.g., ][for a review]{heb09}.
They are located in the HR diagram at the extreme blue extent of the horizontal branch stars; they are sometimes called extreme horizontal branch (EHB) stars. 
Some of the sdB stars show multi-periodic oscillations \citep[see e.g.,][for a review]{fon06}. 
Two groups are known; members of the hotter ($T_{\rm eff} \gtrsim 30,000$K) group pulsate in p-modes with periods of $40-400$~s, while members of the cooler group pulsate in g-modes with periods of $2,000-9,000$~s. 
 Around the boundary of the two groups some hybrid stars are observed. 

Pulsations in sdB stars are excited, as in the B-type main-sequence variables, by the kappa-mechanism at the opacity bump ($T\sim2\times10^5$K) caused mainly by Fe-group elements. 
The relation between the hotter p-mode pulsating sdBs and the cooler g-mode pulsators is similar to that of   the $\beta$ Cephei stars and the SPB stars.
For oscillations in sdB stars to be excited, however, microscopic diffusion plays an essential role.
In the chemically homogeneous envelops with observed heavy element abundances, the kappa-mechanism is too weak to excite pulsations of sdB stars. 
Fe-group metals must be levitated and accumulate in the layers where the kappa-mechanism works \citep{cha01,jef06}. 
In other words, the presence of pulsating sdB stars clearly indicates that microscopic diffusion of heavy elements actually operates in the envelopes of these stars. 
Making theoretical predictions of  the strength of microscopic diffusion is difficult because the diffusion is affected by rotation, winds, etc.
Comparison of observed parameter ranges for the variable sdB stars with theoretical predictions should eventually  provide useful constraints to the theory.
The sdB stars are thought to be in their core-helium burning stage after having lost most of the hydrogen rich envelope. 
However, the detailed evolutionary route for a star to become an sdB star is not yet clear.
Seismological determinations of the total mass and  H-rich envelope mass will prove useful to test  formation scenarios of sdB stars \citep{fon06}.

\subsection{GW Vir stars}
GW Vir (or variable PG1159) stars are very hot ($80,000 {\rm K} \lesssim T_{\rm eff} \lesssim 180,000{\rm K}$) hydrogen-deficient, chemically peculiar pre-white dwarf stars, some of them are associated with a planetary nebula.
The peculiar and diverse surface chemical compositions, mostly  He/C/O, are thought to be produced by nucleosynthesis in the `born-again-AGB' phase during a late-thermal pulse \citep{wer06}.
GW Vir stars pulsate in g-modes with periods ranging from a few hundred  to a few thousand seconds; they are excited by the kappa-mechanism at the opacity peak around $T\sim2\times10^6$K caused by absorption by the L-shell electrons of Fe and  ionizations of C/O/Ne K-shell electrons \citep{opal92}.  
The excitation of  g-modes in PG1159 stars depends on the C/O abundance in the layers around  $T\sim2\times10^6$K. 
Depending on the chemical composition in the envelope, some stars do not pulsate even in the instability region in the HR diagram \citep{qui07}. 
The red edge of GW Vir variables at $T_{\rm eff} \sim 80,000$K arises due to the drainage of C/O by gravitational settling \citep{gau05}. 
The effective temperature at the observed red-edge constrains models of gravitational settling in and wind mass loss from very hot pre-white-dwarf stars.

A large number of frequencies of the prototype star GW Vir (PG1159-035) was obtained from the  12-day nearly continuous observations of the first WET (Whole Earth Telescope) campaign. 
The asteroseismic analysis for the frequencies has yielded important basic information on this star such as mass, rotation rate, etc \citep{win91}.

\section{Effects of Convection}

\subsection{Convection mechanism of excitation}
The g-mode pulsations in ZZ Cet (variable DA white dwarf) stars are excited around the H-ionization zone where the convective energy flux is much larger than the radiative flux.
The convection turn-over time in ZZ Cet stars is around 0.1 s \citep{bri83} much shorter than the observed range of periods of $10^2 - 10^3$s \citep[e.g.][]{fon08}.
Therefore, the state of convection should be adjusted instantaneously at each phase of a g-mode pulsation. 
\citet{bri83} found that the instantaneously adjusting convection zone drives g-modes of ZZ Ceti stars. 
The effect of perturbed radiative flux at the bottom of the convection zone is immediately distributed in the convection zone; i.e., the entropy changes homogeneously in the convective zone.   
Using a quasi-adiabatic analysis, \citet{gol99} showed  that the instantaneous adjustment leads to thermal energy being stored in the convection zone during the compressed phase of pulsation.
This drives pulsation for the same reason as  the kappa-mechanism does, in which energy is gained by blocking radiative flux during the compressed phase  \citep[e.g.][]{cox74}. 
In the ZZ Ceti star case,  the convection zone stores heat during the compressed phase, so that the mechanism should be called {\bf convection mechanism} or {\bf convective driving} \citep{gol99}. 
However, we have to keep it in mind that convection zones realized in hydrodynamic simulations have properties quite different from those based on the mixing-length theory \citep{gau96}.

\subsection{Convection-pulsation coupling}
Except for the ZZ Ceti stars (and probably variable DB stars),  convection time-scales at the bottom of the convection zones  are generally comparable to or longer than the observed pulsation periods.
If convection carries a substantial fraction of total flux, it is necessary to include the effects of pulsation-convection coupling to understand the excitation of the observed pulsations.
Including the coupling is essential to understand the red-edge of the classical instability strip; various authors with different methods have succeeded in reproducing this red-edge, although the reasons for the stabilization do not always agree with each other.
The differences in treating convection-pulsation coupling adopted by different authors are discussed in detail by \citet{hou08}.

Convection-pulsation coupling plays important roles in the excitation of $\gamma$ Dor stars \citep{dup05}, which are high-order g-mode pulsators with periods of around a day.
Before the space observations, especially from the Kepler satellite, ground based observations indicated that $\gamma$ Dor stars were confined to the red part of the $\delta$ Sct instability strip on the HR diagram, which was consistent with the theoretical stability predictions obtained including convection-pulsation coupling \citep{dup05}.
This happy consensus has been overturned by the analysis of early Kepler data \citep{gri10} which shows that all the pulsators are hybrid pulsators having $\delta$ Sct-type  as well as $\gamma$ Dor-type pulsations.
The distribution of $\delta$ Sct-type dominated stars in the instability strip  is similar to that of $\gamma$ Dor-type dominated pulsators.
Furthermore, \citet{gri10} found that there are many Kepler target stars in the instability strip that show no periodic variations to the limit of the error level of the Kepler photometry.
The theory for the convection-pulsation coupling need to be further refined to be consistent with these newly found properties.

\subsection{Blue supergiants}
Several blue supergiants show light variations with periods of a few days indicative of g-mode oscillations \citep{lef07}. 
It may appear strange because g-modes outside of the main-sequence band should be damped due to large radiative damping in the central radiative core as mentioned in the subsection \ref{sec:spb}.
It is found, however, that a convective shell above the hydrogen burning shell in the post main-sequence stage can reflect some g-modes. 
These reflected g-modes do not suffer strong radiative damping, and  hence they are excited \citep{sai06} by the Fe-bump of opacity.  
Such a convective shell in the post main-sequence stage appears if the star is massive enough ($M\gtrsim 12M_\odot$). \citet{god09} suggested that comparing the observed range of the slowly pulsating superginant B stars with theoretical stability analysis for g-modes could constrain the wind mass loss rate and the extent of core-overshooting in the main sequence stage, because these affect the appearance of the convective shell in the post-main-sequence stage. 

\subsection{Oscillatory g$^{-}$ modes in massive stars}
Convective instability in non-rotating non-magnetic stars is monotonic in the adiabatic approximation, which corresponds to the presence of  g$^{-}$ modes with purely imaginary eigenfrequencies. 
Although the physical reason is not clear yet, \citet{shi81} found   g$^{-}$ modes to be overstable (i.e., oscillatory, having a finite real part in the eigenfrequency)  in very nonadiabatic environment. 
The overstable g$^{-}$ modes they found were associated with the convection zone caused by H/He-ionization in the luminous, relatively cool (but hotter than the classical instability strip) models ($10^5 L_\odot,  12 M_\odot$).  
Due to the presence of the Fe-opacity peak at $T\sim2\times10^5$ K, hot massive stars have substantial envelope convection zones \citep{can09}. 
It is predicted that oscillatory g$^{-}$ modes are present in the  hot massive stars, and that the amplitude of the oscillation penetrates into the photosphere \citep{sai11}.
These g$^{-}$ modes may be responsible for the enhanced photospheric turbulence in blue supergiants.  The relation between atmospheric turbulence and the Fe convection zones is extensively discussed by \citet{can09}, while \citet{aer09} consider the turbulence to come from collective effects of pulsations.  

\subsection{Solar-type oscillations in red-giants} 
Thanks to the accurate photometric observations of Kepler and CoRoT satellites, our understanding of the solar-type oscillations in red-giant stars has improved enormously.
The structure of a red-giant is characterized by a dense core of $\sim0.4-0.5 M_\odot$ and a deep convective envelope; the former is a g-mode cavity and the latter a p-mode cavity.
Since the Brunt V\"ais\"al\"a frequency, $N$, is very high in the dense core, every nonradial mode ($\ell > 0$) that might be stochastically excited by convection in the envelope of a red-giant must have a mixed character (g-mode in the core and p-mode in the envelope).
Among the mixed modes for a given $\ell$ there is a series of modes whose amplitude of oscillation is strongly confined in the envelope. They are essentially p-modes;
their frequencies are separated by the same large separation, $\Delta\nu$, determined by the radial modes. 
In other words, mixed modes are strongly confined to the envelope only if the frequencies are close to the hypothetical pure p-mode frequencies in the absence of the core.
Other mixed modes are essentially g-modes which have considerable amplitudes in the dense core, and the frequency spacing is much denser than that of p-modes.
The contribution from the envelope depends on the `distance' from the neighboring p-mode, and it is smallest in the middle of two adjacent p-mode frequencies \citep[see e.g. ][for the properties of nonradial modes in red giants]{dup09}.

As the inertia of p-mode-like modes are much smaller than g-mode-like modes, the amplitudes of the former are larger than the latter modes.
Very accurate Kepler and CoRoT photometry detected not only p-mode-like modes but also some g-mode-like modes (with relatively large contribution from the envelope) around the frequencies of p-mode-like modes, from which g-mode period spacings $\Delta P_{\rm g}$ as well as p-mode large frequency separations $\Delta\nu$ were determined \citep{bed11,mos11}.
Since $\Delta P_{\rm g} \propto [\sqrt{\ell(\ell+1)}\int_0^{r_c}Nd\ln r]^{-1}$ with core radius $r_c$, measuring $\Delta P_{\rm g}$ yields  information on the density of the core.
In fact \citet{bed11} showed that a $\Delta\nu-\Delta P_{\rm g}$ diagram clearly separate  central He-burning clump stars from still ascending (pre-He-ignition) red giants.  
If rotational frequency splittings are detected for both types of modes in a star,  the variation of the rotation frequency in the stellar interior can be deduced.
This would be very useful information for understanding  the angular momentum transport in the stellar interior.

\section{Rotation Effects}
When we observe a nonradial oscillation (of an azimuthal order $m$) in a star rotating with angular frequency $\Omega$, the oscillation frequency we observe differs by $-m\Omega$ from the intrinsic frequency of the star.  
This is just due to the transformation from the rotating to the observer's (inertial) frame of reference (or rotational advection). (We adopt the convention that $m<0$ corresponds to prograde modes.)
Genuine rotation effects influencing stellar pulsations come from two kinds of forces, the Coriolis force $2\boldmath{u\times\Omega}$ and centrifugal force $\varpi\Omega^2$, where $\boldmath{u}$ is the oscillation velocity of matter, and $\varpi$ is the distance from the rotation axis. This section very briefly overviews the rotation effects \citep[for extensive discussions see e.g.,][]{unn89,gou09,aer10}.

The centrifugal force deforms equilibrium stellar structure and hence modifies the oscillation frequencies; the significance of the effect is measured by the ratio of the centrifugal force to the gravitational force $\sim\varpi\Omega^2r^2/(GM)$. 
Thus, the strength of the effect does not directly depend on the oscillation frequency, but p-modes whose oscillation energy is confined to the envelope are more affected than g-modes whose oscillation energy is confined to the core.

The importance of the Coriolis force can be measured by the ratio of the Coriolis force per unit mass 
to the acceleration due to oscillation $(\boldmath{\dot{u}})$; i.e., by $2\Omega/\omega$, where $\omega$ is the angular frequency of oscillation in the co-rotating frame.
Coriolis force can significantly affect low-frequency g-modes even for a slow rotation that causes  negligible centrifugal deformation. 
On the other hand, for high-frequency p-modes the Coriolis force effect is very small even for a rapid rotation.

\subsection{Frequency splittings}
For a slowly (in general differentially) rotating star, Coriolis force and the rotational motion (advection) split a pulsation frequency of a mode of a degree $\ell$  into $2\ell +1$ equally spaced frequencies ($m$-splitting) as
$$\Delta\sigma=-m\int_0^R\Omega(r){\cal K}(r) dr   \quad {\rm where} \quad
{\cal K}={\rho r^2\over J}[\xi_r^2+\ell(\ell+1)\xi_h^2 - 2\xi_r\xi_h-\xi_h^2]
$$
with $J\equiv \int_0^R[\xi_r^2+\ell(\ell+1)\xi_h^2]\rho r^2dr$; $\xi_r$ and $\xi_h$ are radial and horizontal displacements of oscillation, respectively.
The first two terms in ${\cal K}$ are from advection, and the last two terms from the Coriolis force. 
The runs of the kernel ${\cal K}(r)$ in the stellar interior differ from mode to mode; generally, p-modes (g-modes) are sensitive to envelope (core) rotation.
Therefore, we can probe the distribution of the rotation frequency in the stellar interior,  if frequency splittings for more than one mode are observed in a star.
Table 1 lists results of such analyses for some $\beta$ Cephei stars, in which frequency splittings for both low-order p and g-modes are observed.
The core of $\beta$ Cephei stars seem to rotate considerably faster than the envelope.
For relatively rapid rotators,  higher oder terms of ${\cal O}(\Omega^2)$, which break the equality of the spacing, should be included.

For more rapidly rotating stars, the perturbative approach is no longer  applicable and the effects of centrifugal deformation must be accurately included,  based on 2D calculations, to obtain accurate p-mode frequencies \citep{ree06,lov08}.  
In rapidly rotating stars, split frequencies from different modes cross each other,  and differential rotation adds further complexities,  so that mode identification becomes very difficult \citep{deu10}.
Furthermore, in rapidly rotating stars new kinds of pulsation modes (such as chaotic modes) exist, which are absent in the non-rotating stars \citep{lig09}. 
However,  high order ordinary (island) p-modes show some regularities which can be approximated by a new asymptotic formula with a set of new quantum numbers \citep{ree09}. This relation might prove useful for asteroseismology of rapidly rotating stars.

\subsection{g-modes in rapidly rotating stars}
Coriolis force deforms the angular dependence of the amplitude of eigenfunctions from the form of a single spherical harmonic. Generally, amplitude tends to be confined to the equatorial range; for large $\Omega/\omega$, an additional latitudinal nodal line appears for retrograde g-modes and for prograde tesseral modes reducing their visibility. Only prograde sectoral modes among g-modes are  immune to such significant deformation of the latitudinal amplitude distribution  \citep[e.g.,][]{lee90}. 
In addition, nonadiabatic analyses indicate that prograde sectoral g-modes are less affected by damping due to mode couplings \citep{apr11}.  

Many properties of g-modes in rotating stars can be understood with the help of the so called `traditional approximation' in which the horizontal component of angular velocity of rotation, $\Omega\sin\theta$ is neglected. 
Under this approximation, differential equations for nonradial pulsations are the same as those for a nonrotating star except that $\ell(\ell+1)$ is replaced with the eigenvalue $\lambda$ of Laplace's tidal equation. The eigenvalue $\lambda$ depends on $\Omega/\omega$ and approaches $\ell(\ell+1)$ as $\Omega/\omega \rightarrow 0$.  
Thus, in a rotating star $\sqrt{\lambda}$ represents the latitudinal degree of the nonradial pulsation.
For a given $\Omega/\omega$ and $m$, an infinite number of $\lambda$ values are possible for  the latitudinal dependence of eigenfunctions (Hough functions), corresponding to the fact that an infinite number of $\ell$s are allowed for a given $m$ in a non-rotating case.
For retrograde g-modes and prograde tesseral modes $\lambda$ has large values for $\Omega/\omega > 1$, while  for sectoral prograde modes ($\ell =-m$ at $\Omega=0$) $\lambda$ is slightly smaller than $\ell(\ell+1)$ \citep[e.g.,][]{lee97,aer10}.   Thus, the prograde sectoral modes have the special character that the latitudinal dependence is not modified drastically even at large $\Omega/\omega$ values.

Many rapidly rotating Be stars have effective temperatures similar to those of SPB stars (Fig.~1) in which high radial-order g-modes are excited.  
For these Be stars the Fourier spectra of photometric data from space show grouped frequencies \citep{wal05,cam08,nei09}. 
These groupings can be understood as high-order prograde g-modes whose frequencies in the corotating frame $\omega_{\rm g}$ are much smaller than the rotation frequency $\Omega$.
Then, the  frequencies in the observers frame, $\omega_{\rm g}+|m|\Omega$, should be grouped around $\Omega$ and $2\Omega$ corresponding to $m=-1$ and $m=-2$ modes, respectively.
This property can be used to determine the rotation frequencies of rapidly rotating B stars by just looking at the distribution of the pulsation frequencies in the Fourier spectra \citep{cam08}.
Although theoretical models for these g-modes in Be stars agree qualitatively  with the observations, no quantitative agreement is obtained for each g-mode frequency. 
This situation will be improved in the near future with more accurate calculations for g-modes in deformed rotating stars \citep{bal10}.

\subsection{r-modes}
The latitudinal gradient of the Coriolis force generates Rossby-wave type slow global oscillation modes called r-modes; they are retrograde modes with a nearly toroidal  (divergence-free) displacement vector, proportional to $(\nabla Y_\ell^m)\times\vec{r}$.  
The frequency of an r-mode in the co-rotating frame is given as 
$2m\Omega[\ell(\ell+1)]^{-1}+ {\cal O}(\Omega^3)$. 
The last ${\cal O}(\Omega^3)$ term arises from the deviation of the displacement  from the toroidal form; because of the deviation  r-modes can propagate in the radial direction and  have nonadiabatic effects such as driving and damping \citep[e.g.,][]{pro81,sai82}.  
Nonadiabatic analyses predict  r-modes to be excited by the kappa-mechanism at the Fe-opacity bump in rotating B-type stars \citep{tow05,lee06}.  

The r-modes with $\ell=1 (m=1)$ have a special character; that is, the frequencies in the inertial frame are very small because $2m\Omega[\ell(\ell+1)]^{-1}=\Omega$. 
These modes propagate azimuthally opposite to the rotational direction at nearly the same speed as rotation, so that we see almost the same phase of the oscillation for a long time.
Space photometry with the MOST, CoRoT, and Kepler satellites indicates the presence of very low frequency oscillations in rapidly rotating stars \citep[e.g.,][]{wal05,nei09,bal11b}.
If these low-frequency oscillations are confirmed to come from the stars (i.e. not being instrumental), they can be interpreted as $\ell=1$ r-modes.

\section{Pulsations of magnetic stars}
Some chemically peculiar A-F stars (Ap stars) have strong ($1\sim30$kG) global (mostly dipole) magnetic fields. 
The surface chemical anomaly is explained by microscopic diffusion (radiative levitation) in the magnetically stabilized and slowly rotating atmosphere.
Although it is somewhat contradictory to the quiet environment needed for the diffusion, some relatively cool Ap stars show high-order p-mode oscillations with periods of $6 - 21$~min; their velocity amplitudes in the outermost layers can reach a few km/s. 
These variables are called roAp (rapidly oscillating Ap) stars; the first stars of this group were found by \citet{kur82} and presently $\sim40$ members are known \citep{kur06}.  
The high-order p-modes are thought to be excited by the kappa-mechanism at the hydrogen ionization zone where convection is expected to be suppressed  by the strong magnetic field \citep{bal01}.  However, pulsation frequencies of some roAp stars exceed the acoustic cut-off frequency and cannot be excited by the kappa-mechanism \citep[e.g.][]{sai10}. 

With a polar strength of  a few kG, the magnetic energy density surpasses the thermal energy density in the outermost layers including some sub-photospheric layers.
The p-modes in roAp stars, whose kinetic energy is confined to the outer envelope,  are strongly affected by the Lorentz force so that they have peculiar properties not seen in other type stars.
In many cases, observed pulsation amplitudes of a roAp star modulate with rotation phase.
To explain this, \citet{kur82} introduced the oblique pulsator model, in which the pulsation is axisymmetric to the magnetic axis which is inclined to the rotation axis, so that we see different aspects of the pulsation at different rotation phases.
Pulsations are influenced not only by magnetic fields but also by rotation whose axis is inclined to the magnetic axis, so that the pulsation axis cannot be strictly tied to the magnetic axis but should oscillate elliptically \citep{big02}.  
However, this effect is expected to be small in most cases because the magnetic effect dominates the effects of slow rotation.  
The magnetic field not only fixes the pulsation axis but also modifies the distribution of amplitude on the surface; i.e., the angular dependence of the eigenfunction cannot be expressed by a single spherical harmonic \citep[e.g.][]{dzi96,sai04}; this fact is confirmed observationally \citep{koc06}.

The presence of magnetic fields introduces new kinds of waves, in addition to modifying the properties of acoustic waves. 
Fast and slow waves are relevant to axisymmetric roAp pulsations.  
The phase velocities of the fast and slow waves are $(c_{\rm s}^2 + V_{\rm A}^2)^{1/2}$  and  $c_{\rm s}V_{\rm A}(c_{\rm s}^2 + V_{\rm A}^2)^{-1/2}$, respectively, where $c_{\rm s}$ is the sound speed and $V_{\rm A}$ Alfv\'en speed. If $c_{\rm s} \gg V_{\rm A}$ (in deep interior), fast waves and slow waves correspond to p-mode pulsations and magnetic oscillations, respectively, while if $c_{\rm s} \ll V_{\rm A}$ (in outermost layers), slow waves and fast waves correspond to p-modes and magnetic waves, respectively.

Stellar pulsation generates magnetic waves due to coupling in the layers where $c_{\rm s} \sim V_{\rm A}$. The generated waves propagate inward as slow waves whose wavelength becomes shorter and shorter as they propagate deeper interior.  
Therefore, the slow waves are  believed to be dissipated before reaching the stellar center \citep{rob83}; i.e.,  the coupling causes damping of the pulsations.
The strength of the coupling (i.e., damping rate) varies cyclically as a function of $\nu B_{\rm p}^{1/(n-1)}$ \citep{cun00,sai04}, where $\nu$ is the oscillation frequency, $B_{\rm p}$ the polar strength of dipole magnetic field, and $n$ the effective polytropic index.  
At a few kG of magnetic field as typical in roAp stars, the magnetic effect on low-order p-modes is relatively small and the damping rate increases with $B_{\rm p}$.
This effect might explain the absence of low-order ($\delta$ Sct) type oscillations in roAp stars, which are found in the $\delta$ Sct instability strip \citep{sai04,sai05}.  
The absence of low-order p-modes may also be explained by gravitational settling of helium as for Am stars (\S\ref{sec_am}).  However, since we are now getting to know that many Am stars do pulsate, magnetic damping seems to be more appropriate. 

For high-order p-modes appropriate for the frequency range of roAp stars, magnetic effect on a mode changes cyclically  as a function of $B_{\rm p}$. 
At  $B_{\rm p}$ corresponding to a maximum coupling with a particular mode (i.e., at a maximum damping) the oscillation frequency jumps down by a few $\mu$Hz \citep{cun00}, and the angular dependence of the amplitude distribution is so deformed that expansions using spherical harmonics fail to converge \citep{sai04}.

\section{Evolutionary period changes}
The radius and internal structure of a star change as evolution progresses. 
If a star is a pulsating variable, the periods also change with evolution.
Needless to say, measuring such evolutionary period changes is very important because that means we measure the rate of stellar evolution!
Although it may be difficult to measure small changes for main-sequence pulsators, evolutionary period changes have actually been measured for many evolved pulsators.
Here, we discuss briefly some examples.

A small fraction ($\sim1$\%) of Mira variables  (radially pulsating AGB stars) show large (positive or negative) secular period changes: $|d\ln P/dt| \sim (0.5 - 1)\times 10^{-2}$yr$^{-1}$ \citep[e.g.][]{tem05}.
Such rapid period changes are consistent with rapid radius changes that occur during  thermal pulses (or helium-shell flashes) \citep{woo81}.
Only a small fraction of Mira variables show such rapid changes because the duration of a thermal pulse is much shorter than the inter-pulse time interval.

Next fastest evolutionary period changes are observed in the hot hydrogen-deficient star V652 Her which pulsates very regularly with a period of 0.108 d in the radial fundamental mode.
\citet{kil05} obtained a period change rate of $d\ln P/dt = -2.5\times10^{-4}$/year. The rapid rate indicates that the star contracts rapidly, which seems consistent with a rapid contraction after its formation from a double WD merger \citep{sai00}.
A period decrease rate for the similar star BX Cir is also obtained by \citet{kil05} but it is $\sim20$ times slower.

Rates of period changes have been measured for many classical Cepheids. 
Most of them are compatible with theoretical models being in the first,  second, or  third crossing of the instability strip \citep{tur06}.
Generally, longer period cepheids change more rapidly because the stellar mass is larger and hence their evolution is faster; e.g., $|d\ln P/dt| \sim 10^{-7}$ yr$^{-1}$ at $P=4$ d,  $\sim 10^{-6}$yr$^{-1}$ at $P=10$ d, and $\sim 10^{-4.5}$ yr$^{-1}$ at $P=30$~d for the second and third crossings. 
Polaris has a period of 3.97 d which increases at a rate of $d\ln P/dt = 1.4\times10^{-5}$ \citep{spr08} much faster than the rate for the third crossing.
The rate of change tells us that Polaris is most likely on the first crossing, evolving toward the core He-ignition \citep{tur06}.

As a white dwarf cools, the periods of  g-modes increase because the Brunt-V\"ais\"al\"a frequency in the white dwarf envelope decreases as cooling progresses.
The rate of the period change reflects the WD cooling rate which depends on the luminosity, mass, and the composition of the core.
The fastest period change observed among the WDs is $d\ln P/dt = 8.0\times10^{-6}$ yr$^{-6}$ obtained for the hot H-deficient pre-WD GW Vir  \citep{cos08}.
The slowest period change ever obtained is probably $d\ln P/dt =  5.2\times10^{-10}$ of the ZZ Ceti star  G 117-B15A, the rate is consistent with a C/O core \citep{kep05}.

Finally, let me add a few words about expected period changes of pre-main sequence (PMS) $\delta$~Sct stars. They have frequency spectra simpler than those of MS $\delta$ Sct stars, and mode identification is not very diffidult \citep{gue09}.
There are $\sim40$ known PMS $\delta$ Sct stars \citep[e.g.,][]{zwi08} but no evolutionary period change has been measured yet.
The expected rates are $d\ln P/dt\sim-5\times 10^{-7}$yr$^{-1}$ ($2M_\odot$) to $\sim-10^{-6}$yr$^{-1}$ ($2.7M_\odot$), which can be measured in the near future.

\acknowledgements 
I am grateful to Alfred Gautschy for his useful comments.

\bibliography{hsaioref}

\end{document}